%Paper: hep-ph/9306226
%From: My Account <me@cryptons.tamu.edu>
%Date: Fri, 4 Jun 93 11:36:13 -0800
%Date (revised): Fri, 2 Jul 93 08:41:29 -0800

% revised 01JUL93 after change in dilaton scenario parameters
\input harvmac

\def\Titleh#1#2{\nopagenumbers\abstractfont\hsize=\hstitle\rightline{#1}%
\vskip .5in\centerline{\titlefont #2}\abstractfont\vskip .5in\pageno=0}

\def\CTPa{\it Center for Theoretical Physics, Department of Physics,
      Texas A\&M University}
\def\CTPb{\it College Station, TX 77843-4242, USA}
\def\HARCa{\it Astroparticle Physics Group,
Houston Advanced Research Center (HARC)}
\def\HARCb{\it The Woodlands, TX 77381, USA}

\def\CERN{\it CERN Theory Division, 1211 Geneva 23, Switzerland}
\def\ie{\hbox{\it i.e.}}     
\def\eg{\hbox{\it e.g.}}

\def\coeff#1#2{{\textstyle{#1\over #2}}}

\catcode`\@=11 % This allows us to modify PLAIN macros.

\def\lsim{\mathrel{\mathpalette\@versim<}}
\def\gsim{\mathrel{\mathpalette\@versim>}}
\def\@versim#1#2{\vcenter{\offinterlineskip
    \ialign{$\m@th#1\hfil##\hfil$\crcr#2\crcr\sim\crcr } }}
\def\boxit#1{\vbox{\hrule\hbox{\vrule\kern3pt
      \vbox{\kern3pt#1\kern3pt}\kern3pt\vrule}\hrule}}

\def\t1{{\tilde 1}}

\def\JL{J. L. Lopez}
\def\DVN{D. V. Nanopoulos}

\def\GeV{\,{\rm GeV}}
\def\TeV{\,{\rm TeV}}

\def\NPB#1#2#3{Nucl. Phys. B {\bf#1} (19#2) #3}
\def\PLB#1#2#3{Phys. Lett. B {\bf#1} (19#2) #3}

\def\PRD#1#2#3{Phys. Rev. D {\bf#1} (19#2) #3}
\def\PRL#1#2#3{Phys. Rev. Lett. {\bf#1} (19#2) #3}
\def\PRT#1#2#3{Phys. Rep. {\bf#1} (19#2) #3}
\def\MODA#1#2#3{Mod. Phys. Lett. A {\bf#1} (19#2) #3}

\def\TAMU#1{Texas A \& M University preprint CTP-TAMU-#1}

\nref\models{See \eg, {\it String theory in four dimensions}, ed. by M. Dine
(North-Holland, Amsterdam, 1988); {\it Superstring construction}, ed. by
A. N. Schellekens (North-Holland, Amsterdam, 1989).}
\nref\KL{V. Kaplunovsky and J. Louis, \PLB{306}{93}{269}.}
\nref\EN{J. Ellis and \DVN, \PLB{110}{82}{44}.}
\nref\IL{L. Ib\'a\~nez and D. L\"ust, \NPB{382}{92}{305}.}
\nref\Casas{B. de Carlos, J. Casas, and C. Mu\~noz, \NPB{399}{93}{623}
and \PLB{299}{93}{234}.}
\nref\AN{R. Arnowitt and P. Nath, \PRL{69}{92}{725}; P. Nath and R. Arnowitt,
\PLB{287}{92}{89} and \PLB{289}{92}{368}; \JL, \DVN, and H. Pois,
\PRD{47}{93}{2468}; \JL, \DVN, H. Pois, and A. Zichichi, \PLB{299}{93}{262}.}
\nref\LN{For a review see, A. Lahanas and \DVN, \PRT{145}{87}{1}.}
\nref\AEHN{I. Antoniadis, J. Ellis, J. Hagelin, and \DVN, \PLB{194}{87}{231};
J. Ellis, J. Hagelin, S. Kelley, and \DVN, \NPB{311}{88/89}{1}.}
\nref\revamped{I. Antoniadis, J. Ellis, J. Hagelin, and \DVN,
\PLB{231}{89}{65}; \JL\ and \DVN, \PLB{268}{91}{359}.}
\nref\JHreview{For a review see, \JL\ and \DVN, in Proceedings of the 15th
Johns Hopkins Workshop on Current Problems in Particle Theory, August 1991,
ed. by G. Domokos and S. Kovesi-Domokos (World Scientific, Singapore, 1992),
p. 277.}
\nref\LNZb{\JL, \DVN, and A. Zichichi, \TAMU{68/92}, CERN-TH.6667/92, and
CERN-PPE/92-188.}
\nref\LNWZ{\JL, \DVN, X. Wang, and A. Zichichi, \TAMU{76/92} and
CERN-PPE/92-194 (to appear in Phys. Rev. D).}
\nref\LNPWZh{\JL, \DVN, H. Pois, X. Wang, and A. Zichichi, \PLB{306}{93}{73}.}
\nref\LNPWZ{\JL, \DVN, H. Pois, X. Wang, and A. Zichichi, \TAMU{89/92} and
CERN-PPE/93-16.}
\nref\hera{\JL, \DVN, X. Wang, and A. Zichichi, \TAMU{15/93} and
CERN-PPE/93-64.}
\nref\indirect{\JL, \DVN, and G. Park, \TAMU{16/93}
(to appear in Phys. Rev. D); \JL, \DVN, G. Park, H. Pois, and K. Yuan,
\TAMU{19/93}; R. Gandhi, \JL, \DVN, K. Yuan, and A. Zichichi
(in preparation).}
\nref\Ross{See \eg, J. Casas, Z. Lalak, C. Mu\~noz, and G. Ross,
\NPB{347}{90}{243}; A. de la Macorra and G. Ross, OUTP-92-14P.}
\nref\LNY{\JL, \DVN, and K. Yuan, \NPB{399}{93}{654}.}
\nref\BLM{R. Barbieri, J. Louis, and M. Moretti, CERN-TH.6856/93.}
\nref\price{I. Antoniadis, J. Ellis, S. Kelley, and \DVN, \PLB{272}{91}{31};
D. Bailin and A. Love, \PLB{280}{92}{26}.}
\nref\SISM{S. Kelley, \JL, and \DVN, \PLB{278}{92}{140}.}
\nref\aspects{S. Kelley, \JL, \DVN, H. Pois, and K. Yuan, \NPB{398}{93}{3}.}
\nref\ANc{P. Nath and R. Arnowitt, \PLB{289}{92}{368}.}
\nref\ERZ{J. Ellis, G. Ridolfi, and F. Zwirner, \PLB{262}{91}{477}.}
\nref\DNh{M. Drees and M. Nojiri, \PRD{45}{92}{2482}.}
\nref\LNYdmI{\JL, \DVN, K. Yuan, \NPB{370}{92}{445}.}
\nref\KLNPYdm{S. Kelley, \JL, \DVN, H. Pois, and K. Yuan, \PRD{47}{93}{2461}.}
\nref\muproblem{J. E. Kim and H. P. Nilles, \PLB{138}{84}{150} and
\PLB{263}{91}{79}; \JL\ and \DVN, \PLB{251}{90}{73}; E. J. Chun, J. E. Kim,
and H. P Nilles, \NPB{370}{92}{105}.}
\nref\Casasmu{J. Casas and C. Mu\~noz, \PLB{306}{93}{288}.}
\nref\GM{G. Giudice and A. Masiero, \PLB{206}{88}{480}.}
\nref\DL{L. Durand and \JL, \PLB{217}{89}{463}, \PRD{40}{89}{207}.}
\nref\trileptons{P. Nath and R. Arnowitt, \MODA{2}{87}{331}; R. Barbieri,
F. Caravaglios, M. Frigeni, and M. Mangano, \NPB{367}{91}{28};
H. Baer and X. Tata, \PRD{47}{93}{2739}.}

\nfig\I{The first-generation squark and slepton masses as a function of
the gluino mass, for both signs of $\mu$ and $m_t=150\GeV$. The same values
apply to the second generation. The thickness of the lines and their deviation
from linearity are because of the small $\tan\beta$ dependence.}
\nfig\II{The $\tilde\tau_{1,2}$, $\tilde b_{1,2}$, and $\tilde t_{1,2}$ masses
versus the gluino mass for both signs of $\mu$ and $m_t=150\GeV$. The large
variability in the $\tilde\tau_{1,2}$, $\tilde b_{1,2}$ masses is because of
the large $\tan\beta$ dependence in the off-diagonal elements of the
corresponding mass matrices. Note that $m_{\tilde t_1}$ can be light for
a fraction of the parameter space because of the non-negligible value of
the $A$-parameter.}
\nfig\III{The one-loop corrected $h$ and $A$ Higgs masses versus the gluino
mass for both signs of $\mu$ and $m_t=150\GeV$. Representative values of
$\tan\beta$ are indicated.}

%PPE format
%\centerline{EUROPEAN ORGANIZATION FOR NUCLEAR RESEARCH}
%\medskip
%\Titleh{\vbox{\baselineskip12pt
%\hbox{CERN--PPE/93-??}
%\hbox{? June, 1993}
%\hbox{CERN/LAA/93--??}
%\hbox{CERN--TH.6903/93}
%\hbox{CTP--TAMU--31/93}
%\hbox{ACT--11/93}}}
%TH format
\Titleh{\vbox{\baselineskip12pt
\hbox{CERN--TH.6903/93}
\hbox{CTP--TAMU--31/93}
\hbox{ACT--11/93}}}
{\vbox{\centerline{Towards a Unified String Supergravity Model}}}
\centerline{JORGE~L.~LOPEZ$^{(a)(b)}$, D.~V.~NANOPOULOS$^{(a)(b)(c)}$,
and A. ZICHICHI$^{(d)}$}
\smallskip
\centerline{$^{(a)}$\CTPa}
\centerline{\CTPb}
\centerline{$^{(b)}$\HARCa}
\centerline{\HARCb}
\centerline{$^{(c)}$\CERN}
\centerline{$^{(d)}${\it CERN, Geneva, Switzerland}}
\vskip .1in
\centerline{ABSTRACT}
We present a unified string supergravity model based on a string-derived
$SU(5)\times U(1)$ model and a string-inspired supersymmetry breaking scenario
triggered by the $F$-term of the universally present dilaton field. This model
can be described by three parameters: $m_t$, $\tan\beta$, and $m_{\tilde g}$.
We work out the predictions for all sparticle and Higgs masses and discuss the
prospects for their detection at the Tevatron, LEPI,II, and HERA. We find that
the cosmological neutralino relic density is always within current
expectations (\ie, $\Omega_\chi h^2_0\lsim0.9$). We also consider a more
constrained version of this supersymmetry breaking scenario where the $B$-term
is specified. In this case $\tan\beta$ can be determined
($\tan\beta\approx1.4-1.6$) and implies $m_t\lsim155\GeV$ and $m_h\lsim91\GeV$.
Thus, continuing Tevatron top-quark searches and LEPI,II Higgs searches could
probe this restricted scenario completely.
\bigskip
\bigskip
%TH format
{\vbox{\baselineskip12pt\hbox{CERN-TH.6903/93}\hbox{CTP--TAMU--31/93}
\hbox{ACT--11/93}}}
\Date{May, 1993}
%\Date{}

\newsec{Motivation}
The ultimate unification of all particles and interactions has string theory as
the best candidate. If this theory were completely understood, we would be able
to show that string theory is either inconsistent with the low-energy world
or supported by experimental data. Since our present knowledge of string theory
is at best fragmented and certainly incomplete, it is important to consider
models which incorporate as many stringy ingredients as possible. The number of
such models is expected to be large, however, the basic ingredients that such
``string models" should incorporate fall into few categories: (i) gauge group
and matter representations which unify at
a calculable model-dependent string unification scale; (ii) a hidden sector
which becomes strongly interacting at an intermediate scale and triggers
supersymmetry breaking with vanishing vacuum energy and hierarchically small
soft superpersymmetry breaking parameters; (iii) acceptable high-energy
phenomenology, \eg, gauge symmetry breaking to the Standard Model (if needed),
not-too-rapid proton decay, decoupling of intermediate-mass-scale unobserved
matter states, etc.; (iv) radiative electroweak symmetry breaking; (v)
acceptable low-energy phenomenology, \eg, reproduce the observed spectrum of
quark and lepton masses and the quark mixing angles, sparticle
and Higgs masses not in conflict with present experimental bounds,
not-too-large neutralino cosmological relic density, etc.

All the above are to be understood as constraints on potentially realistic
string models. Since some of the above constraints can be independently
satisfied in specific models, the real power of a string model rests in the
successful satisfaction of all these constraints within a single model.

String model-building is at a state of development where large numbers of
models can be constructed using various techniques (so-called formulations)
\models. Such models provide a gauge group and associated set of matter
representations, as well as all interactions in the superpotential, the
K\"ahler potential, and the gauge kinetic function. The effective string
supergravity can then be worked out and thus all the above constraints can in
principle be enforced. In practice this approach has never been followed in its
entirety: sophisticated model-building techniques exist which can produce
models satisfying constraints (i), (iii), (iv) and part of (v); detailed
studies of supersymmetry breaking triggered by gaugino condensation
have been performed for generic hidden sectors; and extensive explorations of
the soft-supersymmetry breaking parameter space satisfying constraints (iii),
(iv), and (v) have been conducted.

It has been recently noted (see Ref. \KL\ and references therein) that much
model-independent information can be obtained about the structure of the
soft supersymmetry breaking parameters in generic string supergravity models.
Supersymmetry breaking can generally be triggered by  non-zero $F$-terms
for any of the moduli fields of the string model or a non-zero $F$-term for the
dilaton field. In the case the former dominate, the scalar masses are generally
not universal, \ie, $m_i=f_im_0$ where $m_0$ is the gravitino mass,
and therefore large flavor-changing-neutral-currents (FCNCs) \EN\ are
potentially dangerous \IL. The gaugino masses arise from the one-loop
contribution to the gauge kinetic function and are thus suppressed
($m_{1/2}\sim(\alpha/4\pi)m_0$) \refs{\IL,\Casas,\KL}. The experimental
constraints on the gaugino masses then force the squark and slepton masses
into the TeV range \Casas. This phenomenologically unappealing situation is
worsen once cosmological constraints are considered: the neutralino relic
density will be large and one would have to fine-tune the parameters to have
the neutralino mass be very near the Higgs and $Z$ resonances. This situation
is not unlike that for the minimal $SU(5)$ supergravity model, where
dimension-five proton decay operators basically demand such a
soft-supersymmetry breaking scenario \AN.

An important exception to the $F$-term moduli induced scenario occurs if
$f_i\equiv0$ and all scalar masses at the unification scale vanish, as is
the case in no-scale supergravity models \LN. With the additional ingredient
of a flipped $SU(5)$ gauge group (to solve the dimension-five proton decay
problem \AEHN\ and be easily derivable from string theory
\refs{\revamped,\JHreview}) all the above problems
are naturally avoided \LNZb, and interesting predictions for direct
\refs{\LNWZ,\LNPWZh,\LNPWZ,\hera} and indirect \indirect\ experimental
detection follow.

Supersymmetry breaking triggered by the dilaton $F$-term yields {\it universal}
soft-supersymmetry gaugino and scalar masses and trilinear interactions \KL\
(in the notation of \eg, Ref. \LN)
\eqn\I{m_0=\coeff{1}{\sqrt{3}}m_{1/2},\qquad A=-m_{1/2}.}
This situation generally does not occur in the gaugino condensation scenario,
although examples exist in the literature \Ross.
In this paper we explore the phenomenological consequences of this
supersymmetry breaking scenario which complements the string-derived flipped
$SU(5)$ model of Ref. \LNY. This scenario has been studied recently also in the
context of the MSSM in Ref. \BLM. In sum, we see basically two unified string
supergravity models emerging as good candidates for  phenomenologically
acceptable string models, both of which include a flipped $SU(5)$ observable
gauge group supplemented by matter representations in order to unify at the
string scale $M_U\sim10^{18}\GeV$ \refs{\price,\SISM}. Supersymmetry breaking
is triggered by either: (a) a moduli $F$-term such that the scalar masses
vanish, or (b) the dilaton $F$-term. In either case, after enforcement of the
above constraints, the low-energy theory can be described in terms of just
three parameters: the top-quark mass ($m_t$), the ratio of Higgs vacuum
expectation values ($\tan\beta$), and the gluino mass ($m_{\tilde g}\propto
m_{1/2}$). Therefore, measurement of only two sparticle or Higgs masses would
determine the remaining thirty. Moreover, if the hidden sector responsible for
these patterns of soft supersy

mmetry breaking is specified, the gravitino mass ($m_0$) will  also be
determined and the supersymmetry breaking sector of the theory will be
completely fixed.

\newsec{Phenomenology}
The procedure to extract the low-energy predictions of the model outlined
above is rather standard by now (see \eg, Ref. \aspects): (a) the bottom-quark
and tau-lepton masses, together with the input values of $m_t$ and $\tan\beta$
are used to determine the respective Yukawa couplings at the electroweak scale;
(b) the gauge and Yukawa couplings are then run up to the unification scale
$M_U=10^{18}\GeV$ taking into account the extra vector-like quark doublet
($\sim10^{12}\GeV$) and singlet ($\sim10^6\GeV$) introduced above
\refs{\SISM,\LNZb}; (c) at the
unification scale the soft-supersymmetry breaking parameters are introduced
(according to Eq. \I) and the scalar masses are then run down to the
electroweak scale; (d) radiative electroweak symmetry breaking is enforced by
minimizing the one-loop effective potential which depends on the whole mass
spectrum, and the values of the Higgs mixing term $|\mu|$ and the bilinear
soft-supersymmetry breaking parameter $B$ are determined from the minimization
conditions; (e) all known phenomenological constraints on the sparticle and
Higgs masses are applied (most importantly the LEP lower bounds on the chargino
and Higgs masses), including the cosmological requirement of not-too-large
neutralino relic density. We have scanned the parameter space for
$m_t=130,150,170\GeV$, $\tan\beta=2\to50$ and $m_{1/2}=50\to500\GeV$. Imposing
the constraint $m_{\tilde g,\tilde q}<1\TeV$ we find $m_{1/2}\lsim465\GeV$ and
$\tan\beta\lsim46$. Relaxing the above condition on $m_{1/2}$ simply allows all
sparticle masses to grow further proportional to $m_{\tilde g}$.

\topinsert
\hrule\smallskip
\noindent{\bf Table I}: The value of the $a_i,b_i$ coefficients appearing in
Eq. (2.2) for $\alpha_3(M_Z)=0.118$. The results apply as well to the
second-generation squark and slepton masses.
\smallskip
\input tables
\thicksize=1.0pt
\centerjust
\begintable
$i$|$a_i$|$b_i$\cr
$\tilde e_L$|$0.406$|$+0.616$\nr
$\tilde e_R$|$0.329$|$+0.818$\nr
$\tilde\nu$|$0.406$|$-1.153$\nr
$\tilde u_L$|$1.027$|$-0.110$\nr
$\tilde u_R$|$0.994$|$-0.015$\nr
$\tilde d_L$|$1.027$|$+0.152$\nr
$\tilde d_R$|$0.989$|$-0.030$\endtable
\smallskip
\hrule\medskip
\endinsert

The neutralino and chargino masses show a correlation observed before in
this class of models \refs{\ANc,\LNZb}, namely
\eqn\II{2m_{\chi^0_1}\approx m_{\chi^0_2}\approx
m_{\chi^\pm_1}\approx0.28m_{\tilde g}.}
The first- and second-generation squark and slepton masses can be worked out
analytically,
\eqn\III{m_i=a_i m_{\tilde g}\left[1+b_i\left({150\over m_{\tilde
g}}\right)^2{\tan^2\beta-1\over\tan^2\beta+1}\right]^{1/2},}
with the $a_i,b_i$ given in Table I. The coefficients have been obtained for
$\alpha_3(M_Z)=0.118$. These masses are plotted in Fig. 1. The thickness and
straightness of the lines shows the small $\tan\beta$ dependence, except for
$\tilde\nu$. The results do not depend on the sign of $\mu$, except to the
extent that some points in parameter space are not allowed for both signs of
$\mu$: the $\mu<0$ lines start-off at larger mass values. Note that $m_{\tilde
e_R}/m_{\tilde e_L}\approx0.81$ and $m_{\tilde u_R,\tilde d_R}<m_{\tilde g}
<m_{\tilde u_L,\tilde d_L}$, with the average first-generation squark mass
$m_{\tilde q}\approx1.01m_{\tilde g}$.
%The novelty in this model is the
%relatively large value of the $A$-parameter which has mainly the effect of
%splitting the $\tilde t$ mass eigenstates considerably.
In Fig. 2 we show
$\tilde\tau_{1,2},\tilde b_{1,2},\tilde t_{1,2}$ for the choice $m_t=150\GeV$.
The large variability on the $\tilde\tau_{1,2}$ and $\tilde b_{1,2}$ masses
is due to the $\tan\beta$-dependence in the off-diagonal element of the
corresponding $2\times2$ mass matrices ($\propto
m_{\tau,b}(A_{\tau,b}+\mu\tan\beta)$). On the the other hand, the off-diagonal
element in the stop-squark mass matrix ($\propto m_t(A_t+\mu/\tan\beta)$) is
rather insensitive to $\tan\beta$ but still effects a large $\tilde t_1-\tilde
t_2$ mass splitting (see Fig. 2) because of the significant $A_t$
contribution. The lowest values of the $\tilde t_1$ mass go up with $m_t$
and can be as low as $m_{\tilde t_1}>88,112,150\,(92,106,150)\GeV$ for
$\mu>0\,(\mu<0)$ and $m_t=130,150,170\GeV$.

The one-loop corrected lightest CP-even ($h$) and CP-odd ($A$) Higgs boson
masses are shown in Fig. 3 as functions of $m_{\tilde g}$ for $m_t=150\GeV$.
Following the methods of Ref. \LNPWZh\ we have determined that the LEP lower
bound on $m_h$ becomes $m_h\gsim60\GeV$, as the figure shows. The largest value
of $m_h$ depends on $m_t$; we find $m^{max}_h\approx107,117,125\GeV$ for
$m_t=130,150,170\GeV$. It is interesting to note that the one-loop corrected
values of $m_h$ for $\tan\beta=2$ are quite dependent on the sign of $\mu$.
This phenomenon can be traced back to the $\tilde t_1-\tilde t_2$ mass
splitting which enhances the dominant $\tilde t$ one-loop corrections to $m_h$
\ERZ, an effect which is usually neglected in phenomenological analyses.
The $\tilde t_{1,2}$ masses for $\tan\beta=2$ and are drawn closer together
than the rest. The opposite effect occurs for $\mu<0$ and therefore the
one-loop correction is larger in this case. The sign-of-$\mu$ dependence
appears in the off-diagonal entries in the $\tilde t$ mass matrix  $\propto
m_t(A_t+\mu/\tan\beta)$, with $A_t<0$ in this case. Clearly only small
$\tan\beta$ matters, and $\mu<0$ enhances the splitting. The $A$-mass grows
fairly linearly with $m_{\tilde g}$ with a $\tan\beta$-dependent slope which
decreases for increasing $\tan\beta$, as shown in Fig. 3. Note that even though
$m_A$ can be fairly light, we always get $m_A>m_h$, in agreement with a general
theorem to this effect in supergravity theories \DNh. This result also implies
that the channel $e^+e^-\to hA$ at LEPI is not kinematically allowed in this
model.

The computation of the neutralino relic density (following the methods of
Refs. \refs{\LNYdmI,\KLNPYdm}) shows that none of the points in parameter
space are constrained by cosmology. In fact, we find $\Omega_\chi
h^2_0\lsim0.9$, which implies that in this model cosmologically interesting
values $\Omega_\chi h^2_0$ occur quite naturally (c.f., the model in Ref.
\LNZb\ where Eq. \I\ is substituted by $m_0=A=0$ and $\Omega_\chi
h^2_0\lsim0.25$ is obtained).

\newsec{A special case}
In our analysis above, the radiative electroweak breaking conditions were used
to determine the magnitude of the Higgs mixing term $\mu$ at the electroweak
scale. This quantity is ensured to remain light as long as the supersymmetry
breaking parameters remain light. In a fundamental theory this parameter should
be calculable and its value used to determine the $Z$-boson mass. From this
point of view it is not clear that the natural value of $\mu$ should be light.
In specific models on can obtain such values by invoking non-renormalizable
interactions \refs{\muproblem,\Casasmu}. Another contribution to this quantity
is generically present in string supergravity models \refs{\GM,\Casasmu,\KL}.
The general case with contributions from both sources has been effectively
dealt with in the previous section. If one assumes that only
supergravity-induced contributions to $\mu$ exist, then it can be shown that
the $B$-parameter at the unification scale is also determined \KL,
\eqn\IV{B(M_U)=2m_0=\coeff{2}{\sqrt{3}}m_{1/2},}
which is to be added to the set of relations in Eq. \I. This new constraint
effectively determines $\tan\beta$ for given $m_t$ and $m_{\tilde g}$ values
and makes this restricted version of the model highly predictive, and even
mostly in conflict with experiment, as we now show.

{}From the outset we note that only solutions with $\mu<0$ exist. This is not
a completely obvious result, but it can be partially understood as follows.
In tree-level approximation, $m^2_A>0\Rightarrow\mu B<0$ at the electroweak
scale. Since $B(M_U)$ is required to be positive and not small, $B(M_Z)$ will
likely be positive also, thus forcing $\mu$ to be negative. A sufficiently
small value of $B(M_U)$ and/or one-loop corrections to $m^2_A$ could alter this
result, although in practice this does not happen. A numerical iterative
procedure allows us to determine the value of $\tan\beta$ which satisfies Eq.
\IV, from the calculated value of $B(M_Z)$. We find that
\eqn\V{\tan\beta\approx1.57-1.63,1.37-1.45,1.38-1.40
\quad{\rm for\ }m_t=130,150,155\GeV}
is required. Since $\tan\beta$ is so small ($m^{tree}_h\approx28-41\GeV$), a
significant one-loop correction to $m_h$ is required to increase it above
$\approx60\GeV$ \LNPWZh. This requires the largest possible top-quark masses
and a not-too-small squark mass. However, perturbative unification imposes
an upper bound on $m_t$ for a given $\tan\beta$ \DL, which in this case
implies \aspects
\eqn\VI{m_t\lsim155\GeV,}
which limits the magnitude of $m_h$
\eqn\VII{m_h\lsim74,87,91\GeV\qquad{\rm for}\qquad m_t=130,150,155\GeV.}
Lower values of $m_t$ are disfavored experimentally.

In Table II we give the range of sparticle masses that are allowed in this
case. Clearly, continuing top-quark searches at the Tevatron and Higgs searches
at LEPI,II should probe this restricted scenario completely.

\topinsert
\hrule\smallskip
\noindent{\bf Table II}: The range of allowed sparticle and
Higgs masses in the restricted supersymmetry breaking scenario discussed in
Sec. 3. The top-quark mass is restricted to be $m_t\lsim155\GeV$. All masses
in GeV.
\smallskip
\input tables
\thicksize=1.0pt
\centerjust
\begintable
$m_t$|$130$|$150$|$155$\cr
$\tilde g$|$335-1000$|$260-1000$|$640-1000$\nr
$\chi^0_1$|$38-140$|$24-140$|$90-140$\nr
$\chi^0_2,\chi^\pm_1$|$75-270$|$50-270$|$170-270$\nr
$\tan\beta$|$1.57-1.63$|$1.37-1.45$|$1.38-1.40$\nr
$h$|$61-74$|$64-87$|$84-91$\nr
$\tilde l$|$110-400$|$90-400$|$210-400$\nr
$\tilde q$|$335-1000$|$260-1000$|$640-1000$\nr
$A,H,H^+$|$>400$|$>400$|$>970$\endtable
\smallskip
\hrule\medskip
\endinsert

\newsec{Prospects for detection and conclusions}
The sparticle and Higgs spectrum shown in Figs. 1,2,3 and Table II can be
explored partially at present and near future collider facilities:
\item{(i)}{\it Tevatron}: The search and eventual discovery of the top quark
will narrow down the three-dimensional parameter space considerably; even
possibly ruling out the restricted scenario discussed in Sec. 3 if
$m_t>150\GeV$. The trilepton signal in $p\bar p\to \chi^0_2\chi^\pm_1X$,
where $\chi^0_2$ and $\chi^\pm_1$ decay leptonically, is a clean test of
supersymmetry \trileptons\ and in particular of this class of models \LNWZ.
We expect that some regions of parameter space with
$m_{\chi^\pm_1}\lsim150\GeV$ could be probed with $100\,{\rm pb}^{-1}$. The
relation $m_{\tilde q}\approx1.01 m_{\tilde g}$ for the $\tilde u_{L,R},\tilde
d_{L,R}$ squark masses should allow to probe the low end of the squark and
gluino allowed mass ranges (at least for $\mu>0$, see Fig. 1). The $\tilde t_1$
mass can be below $100\GeV$ for sufficiently low $m_t$. As the lower bound on
$m_t$ rises, this signal becomes less accessible. The actual reach of the
Tevatron for the above processes depends on its ultimate integrated luminosity,
and will be explored elsewhere.
\item{(ii)}{\it LEPI,II}: The lower bound on the Standard Model Higgs boson
mass could still be pushed up several GeV at LEPI and therefore the strict
scenario of Sec. 3 (which requires $m_h\approx61-91\GeV$) could be further
constrained at LEPI and definitely tested at LEPII. In the general case, at
LEPII only a fraction of the Higgs mass range could be explored, generally for
small $\tan\beta$ values (see Fig. 3). The $e^+e^-\to hA$ channel
will be open only for low $m_{\tilde g}$. Chargino masses
below the kinematical limit ($m_{\chi^\pm_1}\lsim100\GeV$) should not be a
problem \LNPWZ, although $m_{\chi^\pm_1}$ can be as high as $\approx285\GeV$ in
this model. Charged slepton pair production is accessible at LEPII for a small
fraction of the parameter space (see Fig. 2).
\item{(iii)}{\it HERA}: The elastic and deep-inelastic contributions to
$e^-p\to\tilde\nu\chi^-_1$ and $ep\to\tilde e_R\chi^0_1$ \hera, should probe a
non-negligible fraction of the parameter space since both $\tilde\nu$ and
$\tilde e_R$ are light for low $m_{\tilde g}$.
\medskip

We conclude that the well motivated string-inspired/derived model presented
here could soon be probed experimentally, and a strict version of it even be
ruled out. The various ingredients making up our model are likely to be present
in actual fully string-derived models which yield the set of supersymmetry
breaking parameters in Eq. \I. The search for such a model is imperative,
although it may not be an easy task since in traditional gaugino condensation
scenarios Eq. \I\ is usually not reproduced. Moreover, the requirement of
vanishing vacuum energy may be difficult to fulfill, as a model with these
properties and all the other ones outlined in Sec. 1 is yet to be found.

\bigskip
\bigskip
\bigskip
\noindent{\it Acknowledgments}: This work has been supported in part by DOE
grant DE-FG05-91-ER-40633. The work of J.L. has been supported by an SSC
Fellowship. JLL would like to thank J. White for useful discussions.
We thank L. Ib\'a\~nez for pointing out an error in the relation between
$m_0$ and $m_{1/2}$ as derived in Ref. \KL.

\listrefs
\listfigs
\bye